# Atomic scale symmetry and polar nanoclusters in the paraelectric phase of ferroelectric materials

Andreja Bencan[1], Emad Oveisi[2], Sina Hashemizadeh[3,4], Vignaswaran K. Veerapandiyan[5], Takuya Hoshina[6], Tadej Rojac[1], Marco Deluca[5], Goran Drazic[7], Dragan Damjanovic*[,3]

1) Electronic Ceramics Department, Jozef Stefan Institute, 1000 Ljubljana, Slovenia,
2) Interdisciplinary Center for Electron Microscopy, Ecole Polytechnique Fédérale de Lausanne, 1015 Lausane, Switzerland,
3) Group for Ferroelectrics and Functional Oxides, Institute of Materials, Ecole Polytechnique Fédérale de Lausanne, 1015 Lausanne, Switzerland
4) Present address: Foundation for Research on Information Technologies in Society (IT'IS), Zeughausstr. 43, 8004 Zurich, Switzerland
5) Materials Center Leoben Forschung GmbH, Roseggerstrasse 12, 8700 Leoben, Austria
6) School of Materials and Chemical Technology, Tokyo Institute of Technology, Meguro, Tokyo, Japan
7) Department of Materials Chemistry, National Institute of Chemistry, 1000 Ljubljana, Slovenia

Abstract

The nature of the "forbidden" local- and long-range polar order in nominally nonpolar paraelectric phases of ferroelectric materials has been an open question since the discovery of ferroelectricity in oxide perovskites ($ABO_3$). A currently considered model suggests locally correlated displacements of B-site atoms along a subset of <111> cubic directions. Such off-site displacements have been confirmed experimentally, however, being essentially dynamic in nature they cannot account for the static nature of the symmetry-forbidden polarization implied by the macroscopic experiments. Here, in an atomically resolved study by aberration-corrected scanning transmission electron microscopy (STEM) complemented by Raman spectroscopy, we reveal, directly visualize and quantitatively describe static, 2-4 nm large polar nanoclusters in the nominally nonpolar cubic phases of $(Ba,Sr)TiO_3$ and $BaTiO_3$. These results have implications on understanding of the atomic-scale structure of disordered materials, the origin of precursor states in ferroelectrics, and may help answering ambiguities on the dynamic-versus-static nature of nano-sized clusters.

*email: dragan.damjanovic@epfl.ch

Except in the simplest materials, the local symmetry of atomic arrangements is not identical to the nominal crystal symmetry as predicted by phenomenological and first-principles theories or determined experimentally by methods that capture average structure, such as the commonly deployed diffraction techniques.[1,2] Materials that are organized hierarchically respond to external fields with dynamics that depends on the structure and symmetry at all length scales leading to emergent macroscopic behaviours that are nontrivial to interpret.[3,4] One such emergent phenomenon is macroscopic polarization in the nominally nonpolar paraelectric phases of ferroelectric materials.[5,6] It has been proposed that the symmetry-forbidden macroscopic polarization results from a collective arrangement of hypothetical polar nanoclusters, which themselves are not allowed by the symmetry of the paraelectric phase.[5–14] While existence of the polar nanoclusters in the paraelectric phase has been postulated soon after the discovery of ferroelectricity in the first oxide ferroelectric[7], $BaTiO_3$, their presence has not yet been directly visualized and characterised on atomic scale in any classical (nonrelaxor) ferroelectric.

A widely considered model of the polar nanoclusters in perovskites assumes correlated displacements of B-site cations along a subset of <111> cubic directions[9,14]. The model has been supported by scattering, diffraction and NMR experiments.[12,14–16] However, the dynamic nature of the B-site cation off-site displacement (orientation of polar regions stable on the time scale of ~$10^{-9}$-$10^{-12}$ seconds[12,14,17]) does not appear to be consistent with the static nature of the macroscopic polarization[5,10,11,18] observed in the paraelectric phases of classical and technologically important ferroelectrics $(Ba_{1-x}Sr_x)TiO_3$ (BST) and $BaTiO_3$. Excellent dielectric tunability of BST is exploited at microwave frequencies[19,20] while $BaTiO_3$ is the most important capacitor material. In both cases, a proof of the presence of polar nanoclusters and details of their atomic structure would significantly alter interpretation of loss mechanisms in these materials.[20–22]

$(Ba_{1-x}Sr_x)TiO_3$ is a solid solution of two most studied oxide perovskites. The nominal symmetry of their prototypical structure is cubic $Pm\bar{3}m$. While $SrTiO_3$ is an incipient ferroelectric[23], $BaTiO_3$ is a canonical ferroelectric that transforms on cooling into tetragonal ferroelectric phase at the Curie temperature $T_C$=403 K, followed by transitions into an orthorhombic and a rhombohedral ferroelectric phase at lower temperatures.[24] For compositions with x<0.8, BST exhibits the same sequence of phase transitions as $BaTiO_3$ and is cubic at room temperature for x>0.33[25] (see SI1-Fig. SI1). For x>0.9 the material possesses relaxor characteristics.[26] A behaviour consistent with a polar or at least a noncentrosymmetric structure has been observed in the paraelectric phases of BST,[5,10,27] unmodified

BaTiO$_3$[5,9,11,13,18] and in SrTiO$_3$.[28] The origin, composition, size and atomic structure of the putative polar nanoclusters in the paraelectric phase of classical perovskite ferroelectrics cannot be inferred in detail from measurements of macroscopic properties. The absence of dielectric dispersion typical for relaxors in BST for x>0.1[26] suggests that the structure of these polar nanoclusters, their dynamics and effect on macroscopic properties is different from those of polar nanoregions in relaxor ferroelectrics.[29]

In this study using aberration-corrected STEM in combination with energy dispersive x-rays spectroscopy and supported by Raman spectroscopy, we give a direct atomistic evidence for polar nanoclusters in paraelectric BST and BaTiO$_3$, and discuss quantitatively clusters' atomic-scale structure and polar properties. In addition to answering this longstanding question, the results of the study will have an impact on understanding the dielectric tunability and loss mechanisms in the paraelectric phases of ferroelectrics.[20–22]

**Evidence for noncubic nanoclusters and their polar character.** We start by presenting the evidence for nanoclusters with a symmetry lower than the expected cubic in (Ba$_{0.6}$Sr$_{0.4}$)TiO$_3$ (BST6040). STEM measurements (see Methods) were made at ambient temperature where BST6040 is nominally cubic (T$_C$ ≈273 K on cooling, see SI1-Fig. SI1). Figure 1a shows a high-angle annular dark-field (HAADF) Z-contrast image of a BST6040 grain viewed along the [001]$_{pc}$ zone axis (all orientations are indexed as pseudo cubic - pc). Arrows in Fig. 1a illustrate direction and magnitude of displacement of Ti atoms (B-sites) from their expected ideal cubic positions set by the projected centre of the A-site sublattice (for details see SI2). The longest arrows correspond to a displacement of ~15 pm. Groups of atoms with correlated orientation of displacements are highlighted by arrows' colour in Fig. 1b. These groups of atoms form ~ 2-4 nm large clusters. Importantly, our data suggest that a given nano-size cluster is not isolated from neighbouring clusters by the material's cubic matrix (see caption of Fig. 1 a,b); the whole structure rather resembles a "slush-like structure",[30,31] so that clusters interact directly with each other. Note that those regions with small atom displacements in the plane of the image might exhibit larger displacements in the direction perpendicular to the image plane, and thus probably also do not possess the cubic structure. The positions of oxygen atoms are visible in the annular bright-field (ABF) images in Fig. 1c,d. As illustrated in Fig. 1e, oxygen atoms, too, are displaced (Fig. 1c) from ideal positions (Fig. 1d) expected in the cubic perovskite cell. The local structure of BST6040 is therefore nowhere over the examined area ideally cubic, even though the average structure of the material, as determined by powder x-rays diffraction at room temperature, is cubic perovskite (see SI1, Fig. SI1a).

Further evidence for a local symmetry breaking and presence of noncubic nano-size clusters in BST6040 is obtained from the analysis of Raman spectra, as shown in Fig. 1f (see also Methods). The evolution of the Raman spectra suggests three structural regimes in BST6040: (i) A ferroelectric phase from low temperatures up to 288 K, (ii) a cubic paraelectric phase with polar static disorder and ferroelectric-like distortions up to 573 K, and (iii) a cubic paraelectric phase with dynamic disorder above 573 K. Below 268 K the first-order Raman modes dictated by the Raman selection rules for the ferroelectric phase of BST6040 (marked by arrows in Fig 1f) dominate[27]. These modes appear when the long-range ferroelectric order is established in a system at least in one crystallographic direction[32]. From 268 K to 288 K, the intensity of the ferroelectric modes decreases, and two broad modes at ~220 cm$^{-1}$ and ~550 cm$^{-1}$ (marked by asterisks and dotted lines) become evident. These two modes are ascribed to additional Raman scattering activated by (static) broken translational symmetry caused on the short range by lattice defects (i.e., $Sr^{2+}$ substitution). Such static, disorder-activated modes, appear in the vicinity of the first-order modes associated with the vibrations of lighter atoms (i.e., $BO_6$ octahedra).[33,34,35] These modes are different from second-order Raman modes visible in case of dynamic disorder: their scattering mechanism is essentially first-order. In the presence of static disorder such modes coexist with the first-order (zone-centre) phonons and tend to be more evident upon weakening of the latter, as seen in Fig. 1f above 268 K. The disappearance of the long-range ferroelectric phase and the transition to the regime (ii) of static, defect-induced disorder is clearly seen at 288 K by the disappearance of the ferroelectric modes at ~200 cm$^{-1}$ and ~515 cm$^{-1}$, which is in a good correspondence with the Curie temperature of BST6040. From this temperature onwards, the two disorder-activated modes coexist with the first-order mode at 300 cm$^{-1}$ (dashed square in Fig. 1f), which persists in the Raman spectrum up to 573 K. This mode is related to the relative displacement of the A- and B-site sublattices with respect to the oxygen sublattice in $ABO_3$ perovskites[34] and, as it indicates asymmetry within the $BO_6$ octahedra,[36] it is active only in the polar phase. In other words, the mode suggests the presence of a nonzero dipole moment in the lattice up to 573 K (roughly 300 K above the $T_C$). First-order scattering above 288 K, signalled by the presence of the 300 cm$^{-1}$ mode, is the result of the coupling of normal modes with the defect-induced lattice distortion, leading to a scattering contribution from a large part of the Brillouin zone, within a length scale dictated by the static disorder.[37] Such effects were already seen in $KTaO_3$ incipient ferroelectric[37] and even in strained $BaTiO_3$ thin films[38] and were attributed to the presence of static disorder-driven ferroelectric regions in nominally paraelectric phases, and stress-induced ferroelectric phase stabilization, respectively. Static disorder is also at the basis of the broad appearance of the 300 cm$^{-1}$ mode, which indicates coherence of the ferroelectric phase on the short-range (i.e. a few nanometers) rather than on the long-range. Above 573 K, the transition from static disorder to dynamic disorder is evident from the complete disappearance of modes related to the polar state together with the disorder-activated modes, and the concurrent onset of a broader spectral signature. The broad spectrum above 573 K, in fact, is due to second-order Raman scattering reflecting the phonon density of states of BST6040. Such spectral signature is due to the time-average of symmetry-breaking events with random wavevector[32], such as dynamic uncorrelated displacements of the Ti-cation in random directions. The persistence of first-order phonon modes at 300 cm$^{-1}$ above $T_C$, which are only allowed in the polar state along with the disorder-activated modes centred at 220 cm$^{-1}$ and 550 cm$^{-1}$ is a clear evidence that a defect-driven ferroelectric-like distortion is present in the bulk paraelectric phase of BST6040. Hence, the Raman spectra not only confirm the presence of nanoclusters in the paraelectric phase, but, importantly, also indicate that these clusters are polar.

**Quantification of relative atomic displacements.** We next take a closer and quantitative look into the structure of nanoclusters in BST6040 on the atomic scale. To quantify the structural distortion, the relative displacements of atoms [Ti vs (Ba,Sr), O vs (Ba,Sr) and O vs Ti ]) were determined with respect to expected atomic positions in the cubic phase, Fig. 2. The displacements were calculated from positions of individual atom columns extracted from HAADF and ABF images using 2D Gaussian fit as described in Ref.[39] (see also Methods and SI2). For this calculation, we used HAADF and ABF images taken along $[110]_{pc}$ zone axis, as shown in Fig. 2a,b. The effects of sample mistilt and sample thickness on the displacement measurements were considered and taken into account as explained in SI2. A view along $[110]_{pc}$ zone axis is convenient for the quantitative analysis of displacements as it gives access to columns consisting only of Ti atoms and columns containing only those O atoms that lay in the plane of the $O_6$ octahedra, thus providing a more reliable determination of relative displacements of ions with a low atomic number than the view along the $[001]_{pc}$ zone axis.

The displacements determined from experimental HAADF and ABF images of BST6040 are presented in Fig. 2c and Fig. 2d-f, respectively. The examined area (~4x4 nm$^2$) captures roughly one nanocluster. The same information is compiled in the accompanying polar plots, Fig. 2g,h, where the direction of displacements is clearly delineated. The average measured displacements $\bar{r}$ are in the range from ~10 pm (for Ti vs (Ba,Sr)) to ~21 pm (for O vs Ti). There are two key implications of these results. Firstly, Ti vs (Ba, Sr) displacements measured from HAADF and ABF images are non-zero with similar direction and magnitude indicating non-cubic symmetry. In addition, the different magnitudes and directions of $O^{-2}$ anions displacements with respect to those of $(Ba, Sr)^{+2}$ and $Ti^{+4}$ cations measured from ABF image indicate separation of gravity centres for positive and negative charges, i.e., suggesting a non-zero dipole moment in the plane of the image within the examined area. The 2D displacements qualitatively resemble those in simulated orthorhombic $BaTiO_3$ where, unlike in rhombohedral and tetragonal symmetries, O vs Ba displacements are concentrated away from the origin of the coordinate system (see Fig. SI3-1), possibly indicating monoclinic or lower symmetry of the local 3D structure in BST6040 (see SI3, Fig. SI3-1).

**Chemical analysis.** To see whether the polar nano-size clusters in BST6040 originate from segregation of ferroelectric, $BaTiO_3$-rich regions with a $T_C$ above room temperature, we carried out atomic resolution chemical analysis with energy dispersive X-ray spectroscopy (EDXS) (Fig. 3). The chemical heterogeneity is visible, as expected, only on the level of individual atom columns, Fig. 3a-c, but there is no substantial chemical heterogeneity over the larger examined areas of 20x20 $nm^2$, Fig. 3d-f. For a quantitative estimation of the local stoichiometry of the sample, we modelled cubic BST6040 with a random distribution of Ba and Sr (for details see Fig.SI4-1) and compared the intensities of atomic columns in the simulated images with those in experimental images. This analysis shows that there is no a significant departure of Ba/Sr ratio in our samples from the random distribution of Ba and Sr atoms with 60/40 stoichiometry.

That the material is chemically homogeneous on tens of nm scale and the chemical heterogeneity is present only on column-to-column and unit cell-to-unit cell level, as expected for a random Ba/Sr distribution with the overall 60/40 stoichiometry, was further supported by the observation of macroscopic polarity in samples prepared by radically different procedures (see SI1). It can be therefore stated that the polar nanoclusters are property of BST6040 and not a question of the degree of mixing and interdiffusion of Ba and Sr in a given sample, which could be processing dependent.

**Constraints of the 2D analysis.** To explore the constraints of the 2D analysis of STEM data on the determination of the polar nanoclusters' size and relative displacements of atoms, which take place in 3D, we modelled the displacements in a volume consisting of a polar structure embedded within a cubic matrix (details in SI5 and Figure SI5-1). We conclude that the polar displacements in an embedded volume with a thickness up to few nm can be reliably detected by the procedures used in this work. It was found, in fact, that the atomic displacements associated with deeper regions in the sample (> 5-10 nm) have a marginal influence on the displacement orientation viewed in the STEM images (see SI5-2). Furthermore, since the nano-size clusters appear to be randomly distributed, are roughly isotropic in size (Fig. 1) and we do not see on the surface individual nano-size polar clusters with coherent polarization larger than ~4 nm, we can state that the nano-size clusters observed in BNST6040 are roughly 2-4 nm large in three dimensions.

**Polar nanoclusters in cubic BaTiO$_3$.** We next present and discuss the evidence for polar nanoclusters in the unmodified BaTiO$_3$. This part of the investigation was further complicated by the need to heat samples *in situ* above the T$_C$ (≈403 K) to reach the cubic phase (see Methods). At 473 K we observe a region composed of two polar nanoclusters, Fig. 4. Two populations can be identified for each of the three pairs of relative atomic displacements (Ti vs Ba, O vs Ba and O vs Ti), further confirming that our method in determining quantitatively displacements is sufficiently robust to identify and resolve neighbouring nanoclusters with different orientation of displacements and associated polarisation. The interface between the two populations, spreading roughly through the middle of the examined region, defines the two nanoclusters. The O vs Ti displacements for the two populations are colinear but with different magnitudes. One population of O vs Ti displacements is centred around the zero, indicating a low magnitude of displacements' projection in the plane of the image for that nanocluster. The Ti vs Ba and O vs Ba displacements are making roughly an angle of 90° for both populations and are non-zero. To inspect whether these displacements could correspond to the tetragonal, orthorhombic or rhombohedral polar structure of BaTiO$_3$, we have modelled these structures and compared the associated displacements (SI3 and Fig. SI3-1) with the experimental data (Fig. 4). The symmetry of the experimental displacements in the nominally cubic phase are qualitatively different than in any of the polar phases of BaTiO$_3$. The experimental data are thus consistent with the presence of two polar clusters in the examined region, each possessing an overall monoclinic (or lower) symmetry. The low symmetry of the nanoclusters and their size of roughly 2-4 nm are similar to those found in BST6040. The clusters exhibit coherent relative displacements of cations and oxygen, suggesting their overall polar character.

At 573 K, all displacements are smaller than at 473K, but are still non-zero, with the scattering smaller than at 473 K, and the average structure resembling cubic although some tendency toward non-cubic distortion can be discerned in Ti vs Ba relative displacements (see Fig. SI3-3). These data are in agreement with the results obtained by acoustic emission study[40] which indicated a change in the structure of hypothetical polar nanoclusters in the temperature range between 500 K and 550 K.

The Raman spectra of a BaTiO$_3$ sample (Fig. 4h) confirm the conclusions of the STEM atomic-scale data analysis and suggest a behaviour similar to that in BST6040. First-order Raman modes (marked with arrows) associated with the ferroelectric order (e.g. the mode at 300 cm$^{-1}$) are observed at least up to 438 K (i.e., ~40 K above T$_C$), whereas disorder-activated modes (seen here clearly at ~520 cm$^{-1}$ – the disorder-related mode at ~270 cm$^{-1}$ is convoluted with strong first-order modes up to T$_C$) are detected up to 473 K. This temperature thus marks the approximate upper limit of static disorder in BaTiO$_3$, suggesting polar nanoclusters at least up to 438 K. The disorder in unmodified BaTiO$_3$ is probably activated by lattice defects such as oxygen or metal (Ba- and Ti-cation) vacancies.[33,35] Above 473 K, transition to dynamic disorder is signalled by the fully second-order Raman signature. Considering the expected differences in the temperature measurements between *in situ* STEM and Raman measurements, the transition temperatures between the static and the dynamic disorder obtained by the two techniques agree rather well.

**Discussion.** Our atomic scale studies confirm the presence of static, roughly 2-4 nm large polar nanoclusters above T$_C$ in BST6040 (at room temperature) and BaTiO$_3$ (at 473 K). In BST6040, these regions are not associated with Ba segregation.

The model by Comes *et al.*[9] of disordered cubic phase in perovskite oxides, which is based on diffuse X-ray and electron scattering, proposed off-site dynamic displacements of B-site cations along <111> directions. These displacements exhibit short-range correlation forming chains and anti-chains along <001> axes. A random distribution of those chains gives on average nonpolar, cubic symmetry. The length of the polar chains was predicted to be between 4 and 10 nm.[9,41] Such off-site displacements have recently been confirmed experimentally in BST by refinement of atomic positions obtained from neutron scattering, x-ray absorption and electron diffraction data.[42] However, a recent study of BaTiO$_3$ by synchrotron X-ray diffraction indicated that the diffuse scattering can be explained by anharmonic phonons and also stated that polar chains such as those proposed by Comes *et al.* cannot be stabilized for more than few picoseconds because of a too low energy barrier among different off-site positions.[17] Our results add thus an additional layer of complexity to this long-standing problem: we show that the atomic structure of the nano-size polar clusters in cubic BST and BaTiO$_3$ is essentially static, at least on the scale of hundreds of seconds or minutes (time needed to make STEM or carry out macroscopic property measurements[5]). The question can then be posed what stabilizes polar nanoclusters? We propose that the driving force for stabilization of local polarity in BST and BaTiO$_3$ may be provided by local strains within and around clusters; we have indeed experimentally observed evidence of such strains (see SI6, Fig. SI6-1 and SI6-2). In BST the strains may be due to the size difference of Sr and Ba and could be related to coherent atomic displacements from ideal cubic positions. These unit-cell and cluster-level strains may freeze B-cation within a nanocluster in one of the eight split-sites or at least lead to unequal times the cation spends in different offsite positions.[43] Furthermore, it is known that the size difference of Ba and Sr leads to unequal Sr-O and Ba-O distances in BST while off-site displacement of Ti depends on number of Sr cations around it.[42] A similar role could be played in unmodified BaTiO$_3$ by O, Ba and Ti vacancies and possibly reduced Ti$^{+3}$ cations on B perovskite site (note that Ti$^{+3}$ and host Ti$^{+4}$ possess different size[44]). Significantly, the presence of strains disrupting centro-symmetricity has also been clearly demonstrated by the activation of first-order Raman modes in the nominally paraelectric phase of both BST6040 and BaTiO$_3$.

**Methods:**

*Material preparation:*

(Ba$_{1-x}$,Sr$_x$)TiO$_3$ ceramics (x=0 to 1, with steps of 0.1) were synthesized by solid state reaction of BaTiO$_3$ and SrTiO$_3$ precursors. All investigations for this study were made on composition with x=0.4 (BST6040) and x=0 (BaTiO$_3$). BST6040 exhibits T$_C$ at 273 K during cooling, therefore material's nominal structure at room temperature is cubic, centrosymmetric $Pm\bar{3}m$. The grain size of ceramics is larger than 5 μm. Thus, every sample examined by HR-STEM is a single crystal. For more details see Refs. [5], SI1 and SI5.

*Electron Microscopy:*

The BST6040 samples for STEM were prepared by a combination of mechanical polishing, followed by argon ion beam milling to electron transparency. BaTiO$_3$ samples were analysed in the powder form which was grinded, annealed at 900 °C for 2h in air and slowly cooled down to RT to eliminate residual strains. STEM imaging and energy-dispersive X-ray spectroscopy were performed on a double Cs-corrected FEI Titan Themis 60-300 (equipped with Super-X EDX system) and a probe Cs-corrected Jeol ARM 200 CF (equipped with Centurio EDX system). Thickness of the STEM samples was estimated from low-loss EELS spectra collected by a Gatan Quantum ER Dual EELS spectrometer.

HAADF and ABF images were simultaneously acquired in the form of image series that underwent image alignment procedure to produce an averaged image with reduced statistical image noise (see SI2 for more details).

The central position associated with each atomic column was localized on the averaged HAADF and ABF images using a two-dimensional Gaussian fitting procedure, and displacements for the B- and O-site columns were determined by measuring their displacements from ideal cubic positions. For experimental details of imaging and atom spacing measurement see SI2.

*In situ* heating experiments of $BaTiO_3$ were performed using heating TEM holder made by Protochips (Fusion model). Images were taken at RT, 473 K and 573 K.

STEM image simulations were performed using quantitative image simulation code (QSTEM)[45] with multislice method and frozen phonon approximation. Details of calculations are described in SI2 and SI6 for BST, and in SI3 for BT.

*Raman spectroscopy*

Raman measurements were carried out in a LabRAM 300 spectrometer (Horiba Jobin Yvon, Villeneuve d'Ascq, France) using an Nd:YAG solid state laser with a wavelength of 532 nm in a backscattering geometry. The laser light was focused on the sample surface by means of a long working distance 100x objective (with NA 0.8, LMPlan FI, Olympus, Tokyo, Japan). The effective power at the sample surface was kept below 3 mW. The spectra were collected with a Peltier-cooled Charge Coupled Device (CCD) and visualized in commercial software environment (Origin 2018b, OriginLab Corp., Northampton MA, USA) after correcting for the Bose-Einstein population factor. Temperature dependent Raman measurements were carried out in a Linkam (THMS600, Linkam, Tadworth, UK) temperature-controlled stage.

**Acknowledgements**

A.B, T.R. and G.D. acknowledge funding from the Slovenian Research Agency within programmes P2-0105 and P2-0393 and project J2-2497. Brigita Kmet and Daniele Laub are acknowledged for STEM sample preparation. E.O. thanks Duncan T.L. Alexander and Sorin Lazar for constructive discussions. S.H. and D.D. acknowledge financial support from the Swiss National Science Foundation (No. 200021-159603) and Alberto Biancoli for discussions and help in sample preparations. V. K. V. and M. D. acknowledge support from the Austrian Science Fund (FWF) Project P29563-N36. M. D. acknowledges funding from the European Research Council (ERC) under the European Union's Horizon 2020 research and innovation programme (grant agreement No 817190).

**Author Contributions**

S.H. prepared samples, measured and interpreted macroscopic properties. T.H. prepared samples using different techniques and investigated effects of defects and microstructure on the forbidden polar properties. V.K.V. and M.D. performed Raman measurements and analysis. T.R. critically commented the results, the data analysis and the text. E.O., A.B. and G.D. performed all atomic resolution investigations and analysed and interpreted STEM data. A.B., G.D., E.O., V.K.V., M.D. and D.D. wrote the manuscript with input from all coauthors. D.D. initiated the study and developed the concept.

**Competing Interests statement**

The authors declare no competing financial interests.

**Figures:**

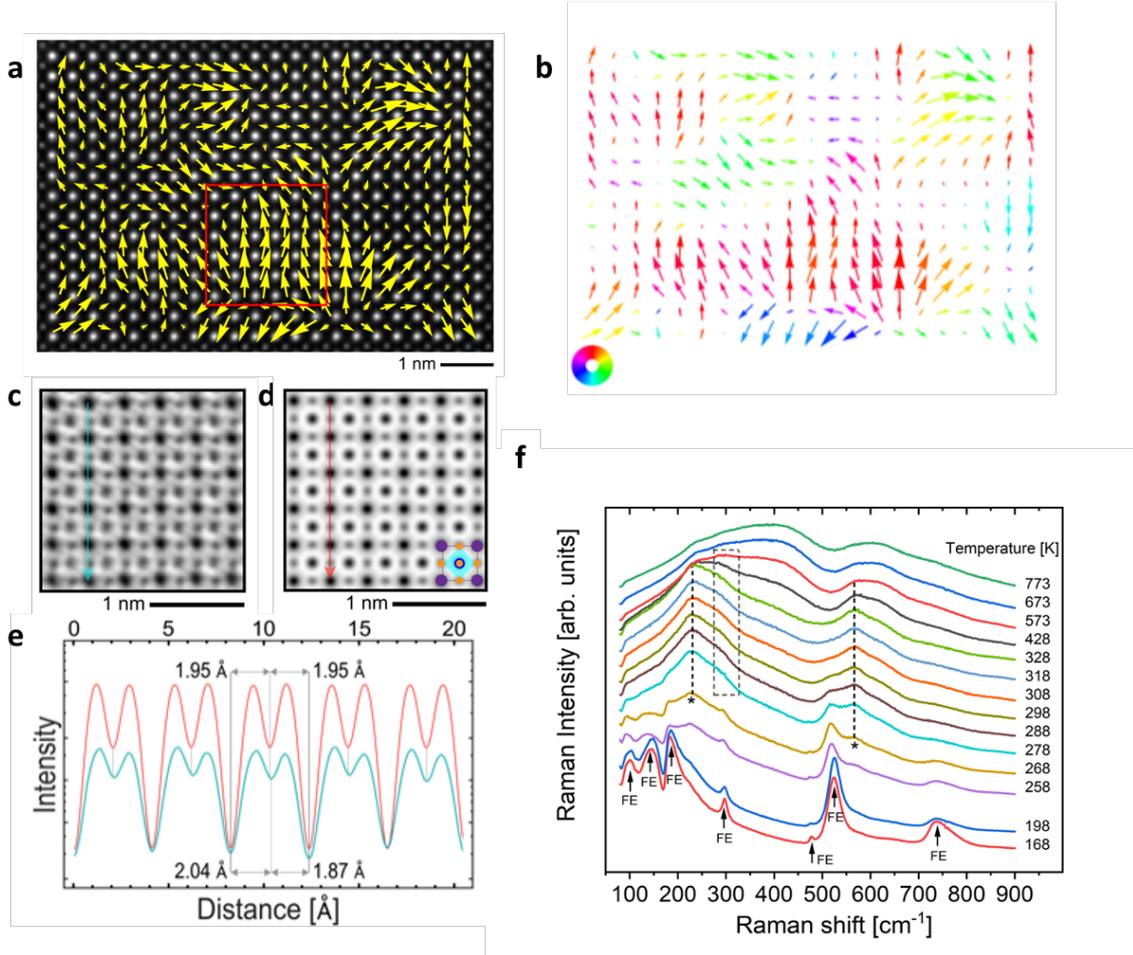

**Figure 1**. **Evidence of polar nanoclusters in cubic BST6040**. **a,** HAADF image of a 14×21 unit-cell area viewed along the $[001]_{pc}$ zone axis. The bright spots correspond to the (Ba,Sr) columns (A-site) and smaller, paler spots to the Ti columns (B-site). Arrows show displacement of B-site atoms measured as the deviation of experimental position of B-sites from the centre of the four neighbouring experimental A-site positions (see SI2). The length and orientation of arrows indicate magnitude and angle of the relative displacement of Ti from the ideal position. **b,** The same as in **a** with the colour of arrows indicating regions of coherent displacements, highlighting different nanoclusters. **c,** Experimental ABF image from the region marked by the red rectangle in **a**. **d,** simulated ABF image of cubic BST6040. **e,** Intensity profiles along the arrows shown in **c** and **d**: cyan, experimental ABF image; red, simulated ABF image with cubic phase; respectively. Displacement of oxygen atoms from the ideal cubic sites are visible in the cyan profile. **f,** Raman spectra of BST6040 as a function of temperature. The modes marked by arrows are characteristic of the ferroelectric phase (FE) and those marked by asterisks indicate broken translational symmetry caused, on the short range, by lattice defects (i.e. $Sr^{2+}$ substitution). The first-order phonon mode at 300 cm$^{-1}$ (enclosed by the dashed square) indicates that the ferroelectric-like distortion is present on the short-range scale well within the paraelectric phase (above $T_C$=273 K). The presence of polar and disorder-activated modes up to 573 K indicates that transition to purely dynamic disorder is attained only above this temperature (see text for details).

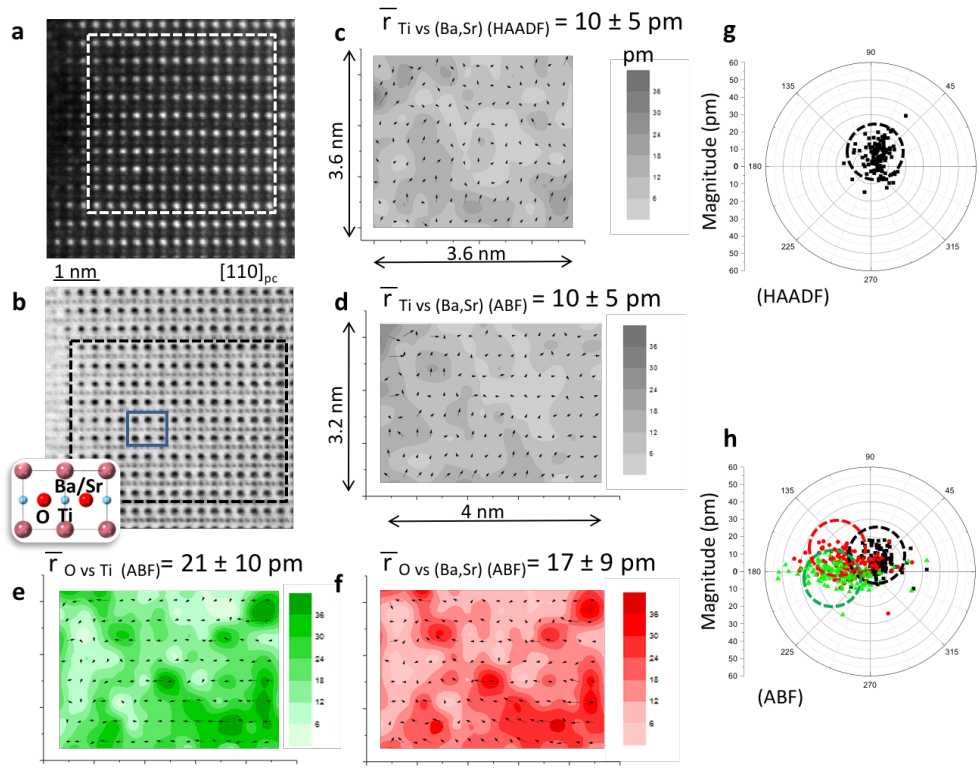

**Figure 2. Analysis of atomic displacements in BST6040. a**, HAADF image and **b,** ABF image, viewed along the $[110]_{pc}$ zone axis of a 15 nm thick BST6040 grain. Dashed rectangles mark areas where displacements were determined. The inset in **b** shows the perovskite unit cell viewed along $[110]_{pc}$ axis. **c-f,** Displacements of Ti vs (Ba, Sr), O vs (Ba, Sr), and O vs Ti determined from HAADF (**c**) and ABF (**d-f**) images with respect to positions of these atoms in an ideal cubic phase with the corresponding average displacements ($\bar{r}$). Arrows represent the direction and magnitude of the displacements and colours represent magnitude of displacement given as contour plots (see colour scale on the right of each image). **g-h**, Compilation of the data presented in **c** - **f** in the form of polar plots, which delineate the orientation of atom displacements. The polar plots show that the symmetry of the local atomic structure is not cubic, is most likely polar and possibly possesses monoclinic or lower symmetry. Dashed circles mark areas where majority of the individual type of displacements are present.

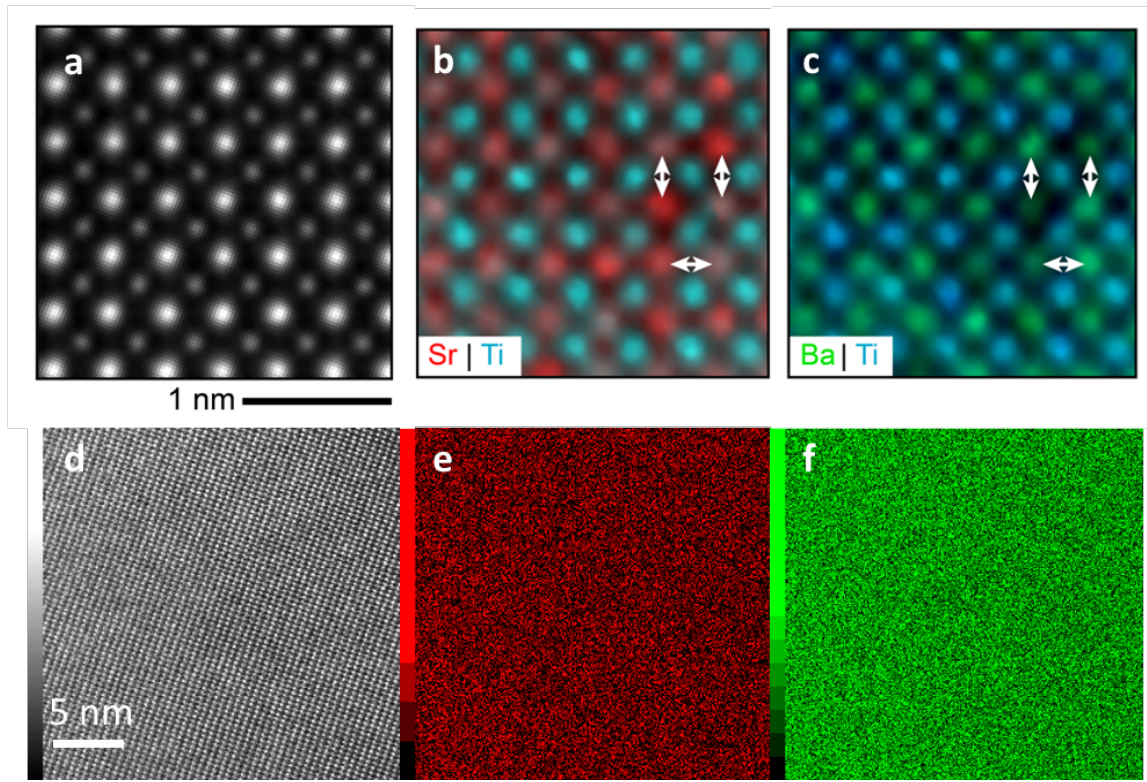

Fig. 3. **Nanoscale and atomically resolved chemical analysis of BST6040. a**, HAADF-STEM image viewed along the [001]$_{pc}$ zone axis of a ~2 nm region from Fig.1a. The corresponding EDXS intensity maps of **b**, Sr/Ti and **c,** Ba/Ti. The arrows indicate Sr (**b**) and Ba (**c**) rich (high intensity) and poor (low intensity) columns. **d,** HAADF-STEM image of a sample segment about 20x20 nm$^2$ in lateral dimensions and corresponding EDXS intensity maps of **e**, Sr-L and **f**, Ba-L lines. The images illustrate chemical homogeneity on several nm scale and heterogeneity on unit cell and column-to-column scale, while the Ba/Sr 60/40 stoichiometry is preserved on all levels (see SI4).

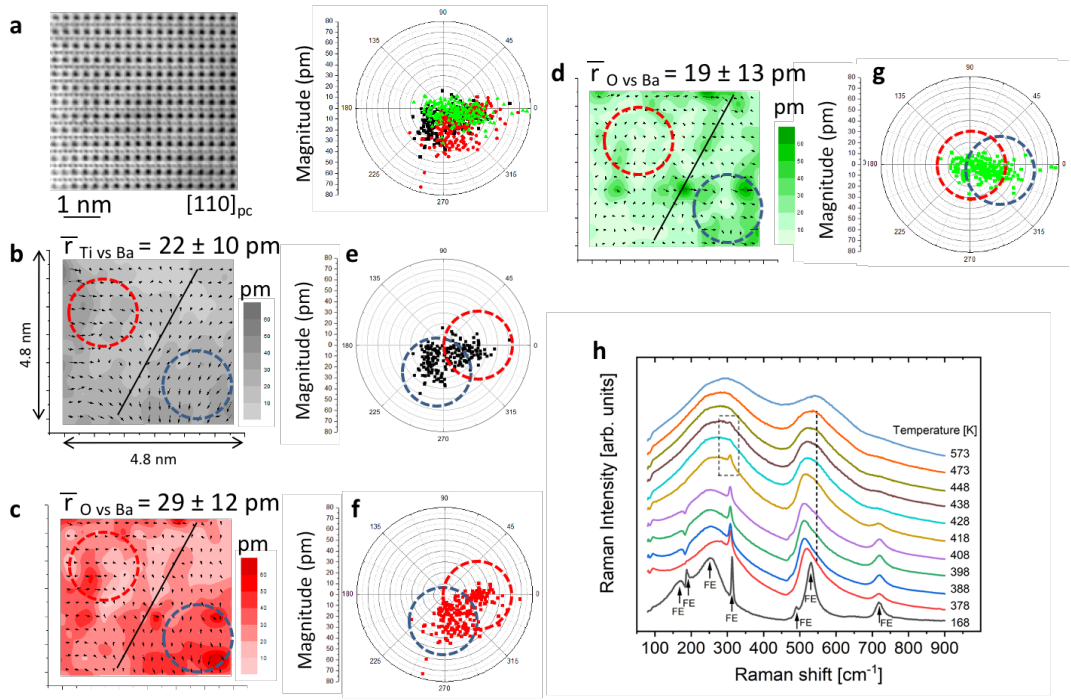

Figure 4. **Atomic displacements and Raman analysis of BaTiO$_3$ at 473 K. a**, ABF image of cubic phase of BaTiO$_3$ viewed along the [110]$_{pc}$ zone axis that was analysed to determine the displacements with composite polar plot. **b**, Ti vs Ba, **c**, O vs Ba and **d**, O vs Ti relative atomic displacements in a BaTiO$_3$ grain. Arrows represent the direction and magnitude of the displacements (see colour scale on the right of each image). **e** - **g**, The same data as in **b-d** presented in polar plots, delineating the orientation of the atomic displacements. The interface between the two polar clusters is indicated by the diagonal lines. The red and blue circles indicate representative areas of each cluster. **h**, Raman spectra of BaTiO$_3$ as a function of temperature. First-order phonon modes associated with the ferroelectric phase are indicated by arrows and (e.g. the mode at 300 cm$^{-1}$, marked by dashed square) are identified up to 438 K (i.e., above T$_C$ of ≈403 K). From 438 K to 473 K, the modes belonging to the static disorder are still present (marked by dotted line). Above 473 K only second-order modes resembling the phonon density of states of BaTiO$_3$ are recorded. Hence, static (polar) disorder is present not only in BST6040 but also in BaTiO$_3$, up to at least 438 K and a transition to the dynamic (non-polar) disorder occurs above 473 K.

# Supplementary Information

# Atomic scale symmetry and polar nanoclusters in the paraelectric phase of ferroelectric materials

Andreja Bencan[1], Emad Oveisi[2], Sina Hashemizadeh[3,4], Vignaswaran K. Veerapandiyan[5], Takuya Hoshina[6], Tadej Rojac[1], Marco Deluca[5], Goran Drazic[7], Dragan Damjanovic*[,3]

1) Electronic Ceramics Department, Jozef Stefan Institute, 1000 Ljubljana, Slovenia,

2) Interdisciplinary Center for Electron Microscopy, Ecole Polytechnique Fédérale de Lausanne, 1015 Lausane, Switzerland,

3) Group for Ferroelectrics and Functional Oxides, Institute of Materials, Ecole Polytechnique Fédérale de Lausanne, 1015 Lausanne, Switzerland

4) Present address: Foundation for Research on Information Technologies in Society (IT'IS), Zeughausstr. 43, 8004 Zurich, Switzerland

5) Materials Center Leoben Forschung GmbH, Roseggerstrasse 12, 8700 Leoben, Austria

6) School of Materials and Chemical Technology, Tokyo Institute of Technology, Meguro, Tokyo, Japan

7) Department of Materials Chemistry, National Institute of Chemistry, 1000 Ljubljana, Slovenia

SI1: Materials preparation and average structure

SI2: Experimental details of HAADF/ABF imaging, (Ba, Sr)TiO$_3$ simulations and displacement measurements

SI3: BaTiO$_3$ simulations and displacement measurements

SI4: Quantitative HAADF STEM analysis of atom column intensities and average chemical composition of (Ba,Sr)TiO$_3$

SI5: Influence of overlapping polar clusters along the viewing direction on displacement measurements

SI6: Evidence of strain associated with polar nanoclusters

**SI 1 : Material preparation and average structure.**

For more details on samples preparation see refs.[1,2] which present results of detailed microstructural characterizations of samples, including by scanning electron microscopy. The X-ray diffraction spectra, Fig. S1-1a, show that the average macroscopic structure of $(Ba_{0.6}Sr_{0.4})TiO_3$ (BST6040) is cubic perovskite at room temperature. The corresponding data for $BaTiO_3$, which possesses tetragonal perovskite structure at room temperature, can be found in Ref.[2] The dielectric permittivity of BST6040 as a function of temperature, Fig. S1-1b, exhibits expected sequence of phase transitions[3] while the thermal hysteresis indicates a first order character of all phase transitions. We have prepared several more compositions in the $(Ba_{1-x},Sr_x)TiO_3$ solid solutions and the trend of the Curie temperature versus composition follows Vegard's law.[1] All these data corroborate microscopic results that Ba- and Sr-titanate precursors are well mixed and that Ba and Sr are homogeneously distributed over A-sites of the perovskite structure.

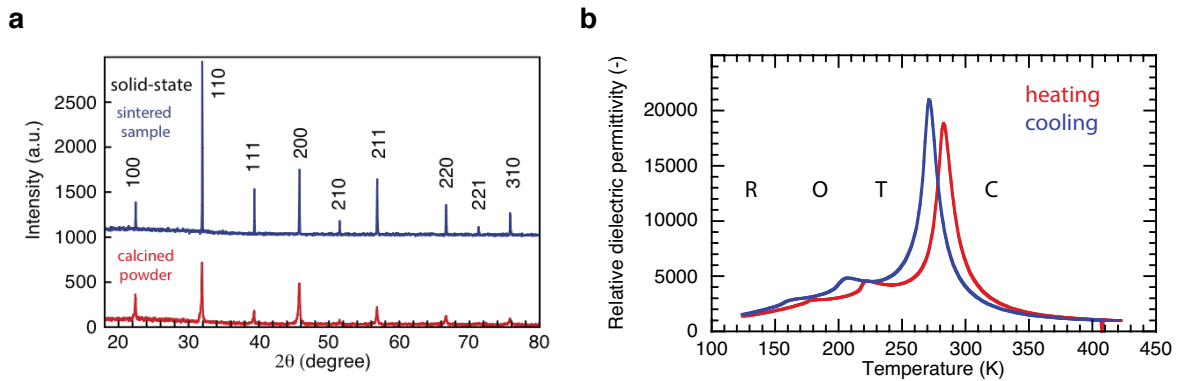

**Fig. SI1-1 Average macroscopic structure**. **a,** X-ray diffraction spectrum of $(Ba_{0.6}Sr_{0.4})TiO_3$ powder and sintered sample, measured at room temperature, showing macroscopic perovskite cubic structure. **b,** Dielectric permittivity of BST6040 as a function of temperature. C, T, O and R indicate, respectively, cubic, tetragonal, orthorhombic and rhombohedral phase.

The macroscopic polar behaviour in the paraelectric phase of $(Ba_{1-x}Sr_x)TiO_3$ and $BaTiO_3$ is a manifestation of presence of nano-size polar clusters. We have observed and systematically studied macroscopic polarity in dozens of $(Ba_{1-x}Sr_xTiO)_3$ samples, including end members, prepared from different precursors (e.g., sol-gel, carbonates, oxides and titanates), by different sintering techniques (e.g., spark plasma sintering and conventional sintering), in different forms (ceramics, tape casted thin layers), with different grain size and with intentionally created nonstoichiometry. The macroscopic polarity was also observed in undoped single crystals of $BaTiO_3$,[1] proving that the macrioscopic polarity and thus polar nanoclusters are a property of these materials and not specific samples. Majority of the samples were prepared at EPFL (Alberto Biancoli and Sina Hashemizadeh) and some at Tokyo Institute of Technology (Prof. Takuya Hoshina). Spark Plasma Sintered samples were obtained from L'Institut de Chimie de la Matière Condensée de Bordeaux (Prof. Catherine Elissalde), and CentralSupélec (Prof. P-E. Janolin). All samples used in this study were prepared at EPFL and are representative samples of this general behaviour.

**SI2: Experimental details of HAADF/ABF imaging, EDXS analysis, (Ba, Sr)TiO$_3$ simulations and displacement measurements**

**The method for displacement calculation**. STEM imaging along the [110] direction was carried out on a Jeol ARM 200 CF operated at 200 kV. HAADF and ABF detectors were used simultaneously at 68–180 and 10–16 mrad collection semi angles, respectively. Beam convergence semi-angle was 24 mrad. To minimise the influence of the specimen drift and scanning irregularities on the atomic column positions HAADF and ABF images were taken as a stack of 20 images; each frame was taken with pixel time of 1.6 μs (2s per frame) using DigiScan Stack Acquisition Tool Digital Micrograph script by Bernhard Schaffer.[4] After the acquisition all images in the stack were aligned using cross-correlation and averaged to obtain low-noise, good quality STEM images using Stack Alignment Digital Micrograph script written by D.R.G. Mitchell.[5]

HAADF and ABF images used for Ti vs (Ba, Sr), O vs (Ba, Sr) and O vs Ti displacement measurements were taken in [110]$_{pc}$ zone axis, represented in terms of pseudo cubic parameters (see Figure SI2-1). Coordinates of atomic column positions were determined using 2D Gaussian fit.

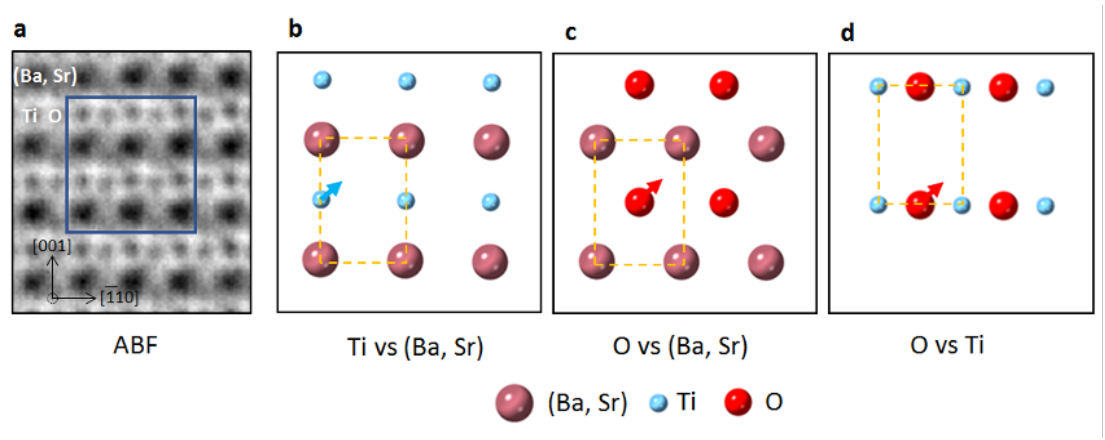

**Figure SI2-1 Atomic displacement calculation. a,** ABF image of (Ba,Sr)TiO$_3$ in [110]$_{pc}$ zone axis where Ba/Sr (most intense large dark circles), Ti (directly below/above (Ba, Sr)) and O (diagonally from (Ba, Sr)) columns are seen. Blue rectangle indicates the area where the methodology of atom columns displacements is explained; **b**, **c** and **d** represent schematics of how individual atom columns displacements were measured. The reference frame was set as indicated by orange dashed rectangle (positions in (Ba, Sr)) and Ti sub-lattices) and the displacements were measured as a deviation from the ideal cubic positions (in the case of Ti vs (Ba, Sr) this position is the middle point at the line between two (Ba, Sr) columns, in the case of O vs Ba this position is the middle point inside the rectangle forming four (Ba, Sr) columns and in the case of O vs Ti this positions is the middle point at the line between two Ti columns). The measured atom column displacements were defined with two parameters, the size of the displacements in pm and the angle of the displacements (0° –360°).

STEM imaging along the [001] direction was carried out on a Thermo Scientific Titan Themis 60-300 at 300 kV. HAADF and ABF images were simultaneously acquired at 90–170 and 9–18 mrad collection semi angles, respectively. In order to reduce statistical image noise and correct linear and non-linear scan distortions, each presented image is the average of a series of rapidly acquired images (approximately 100 frames with 512x512 pixels with 23.5 pm pixel size and a frame time of 500 msec) that underwent rigid and non-rigid alignment using the Smart Align software.[6]

The central position associated with each atomic column was identified on the aligned STEM images by using a two-dimensional Gaussian fitting in the StatSTEM software package.[7] Displacements for the B-site columns were determined by measuring their displacement relative to the mass centre of the four nearest neighboring A-site columns.

Atomic resolution energy-dispersive X-ray spectroscopy was performed on the same machine using a Super-X EDX system comprising four silicon drift detectors, and Velox acquisition software. Spectrum images with 256x256 pixels (pixel size 16.6 pm) and 100 ms frame time (100 frames) were acquired. Elemental maps were acquired using the Ba-Lα, Sr-Lα, and Ti-Kα signals. A combination of Gaussian and Weiner filtering was applied on the elemental maps presented in Figure 3 b,c.

**Effect of sample mistilt and thickness on displacement measurements, (Ba, Sr)TiO$_3$ simulations, and determination of the error in the relative atomic displacement calculations**. The effect of sample mistilt and thickness on atom displacement measurements from ABF images has been reported in the literature.[8–10] Having made special effort to tilt the sample in the exact zone axis, EELS was used to determine the sample thickness. Nevertheless, there is a possibility that due to thermal effects the sample bends to some extent. Below we explain in details our methodology to test for potential mistilt of samples.

We simulated HAADF and ABF images of BST6040 using quantitative image simulation code (QSTEM)[11] with a multi-slice method and frozen phonon approximation. For simulations we used the same instrumental parameters as for experimental imaging (0 defocus, 24 mrad beam convergence semi - angle, 200 kV acceleration voltage). Thermal diffuse scattering (TDS) was included in simulations, so 7-30 repetitions of calculations per one image were used where the atom positions were varied in the interval set by estimated Debye-Waller factors for each calculation.

In order to examine the influence of the sample mistilt and thickness on displacement measurements we performed simulations of HAADF and ABF images of *Pm-3m* cubic structure (ICSD #90006) in [110]$_{pc}$ zone axis for different thicknesses and for a given mistilt angle. Using 2D Gaussian fit we extracted the exact (Ba, Sr), Ti and O atom column positions and calculated Ti vs (Ba, Sr), O vs (Ba, Sr) and O vs Ti displacements. (See Figs. SI2-2 to SI2-4 and Table SI2-1).

As shown in Figure SI2-2 the influence of the sample thickness up to 25nm on the measured displacements of Ti vs (Ba, Sr) in *Pm-3m* BST6040 at zero mistilt is negligible regardless of whether HAADF or ABF images were used. The maximum displacement up to 2 pm was obtained (see Fig. SI2-2 and Table SI2-1). The expected relative atomic displacements for the ideal cubic structure at zero tilt should be zero. Thus, the maximum displacement measured for the simulated cubic structure (~ 2 pm) defines the method's error.

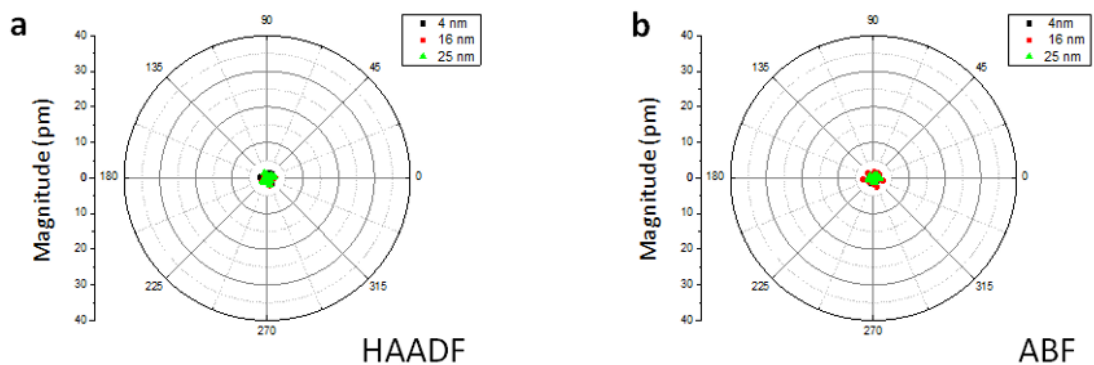

**Figure SI2-2:** Polar figures of Ti vs (Ba, Sr) displacements in P*m-3m* BST6040 measured from simulated **a,** HAADF and **b,** ABF images of cubic symmetry at different sample thicknesses (4,16 and 25 nm), at mistilt angle 0°.

The sample mistilt (0.5°) does not influence Ti vs (Ba, Sr) displacements measured from HAADF images at different thicknesses as shown in Fig SI2-3a. However, the sample mistilt strongly influences both magnitude and direction of displacements measured from ABF images across the examined thickness range as shown in Fig SI2-3b.

In Table SI2-1 average Ti vs (Ba, Sr) displacements at 0 ° and 0.5 ° mistilt measured from simulated HAADF and ABF images of cubic symmetry at different thicknesses are summarized.

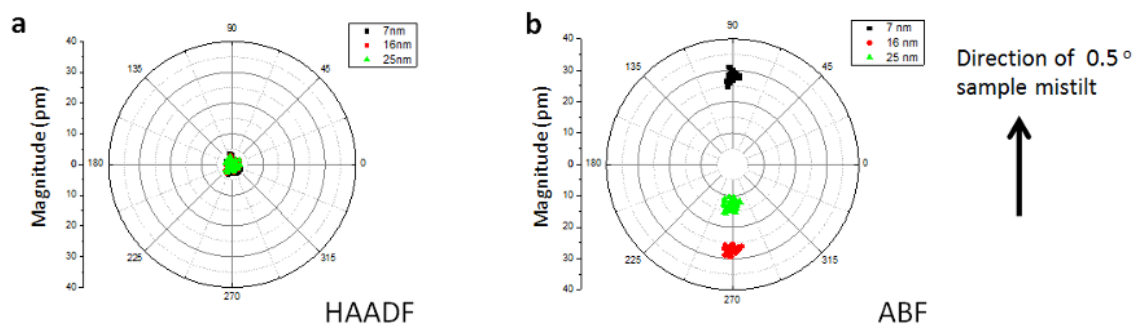

**Figure SI2-3:** Polar figures of Ti vs (Ba, Sr) displacements in Pm-3m BST6040 measured from simulated **a,** HAADF and **b,** ABF images at different sample thicknesses (7,16, 25 nm), at mistilt angle of 0.5°.

**Table SI2-1:** Average Ti vs (Ba, Sr) displacements with one standard deviation, at 0 º and 0.5 º mistilt measured from simulated HAADF and ABF images of cubic symmetry, at different thicknesses.

| Sample thickness/mistilit (nm / º) | Displacements Ti vs (Ba, Sr) from HAADF image (pm) | Displacements Ti vs (Ba, Sr) from ABF image (pm) |
|---|---|---|
| 4/0 | 0.9 ± 0.4 | 0.9 ± 0.4 |
| 16/0 | 1.0 ± 0.5 | 1.3 ± 0.6 |
| 25/0 | 1.0 ± 0.5 | 0.9 ± 0.4 |
| 7/0.5 | 2.0 ± 0.9 | 28.0 ± 1.3 |
| 16/0.5 | 1.6 ± 0.9 | 27.2 ± 1.0 |
| 25/0.5 | 1.6 ± 0.8 | 13.1 ± 1.3 |

Importantly, in the case of 0.5 º mistilt the amplitude and direction of all interatomic displacements (Ti vs (Ba, Sr), O vs (Ba, Sr) and O vs Ti) in *Pm-3m* BST6040 determined from simulated ABF images strongly depend on sample thickness, as can be seen from Fig. SI2-4.

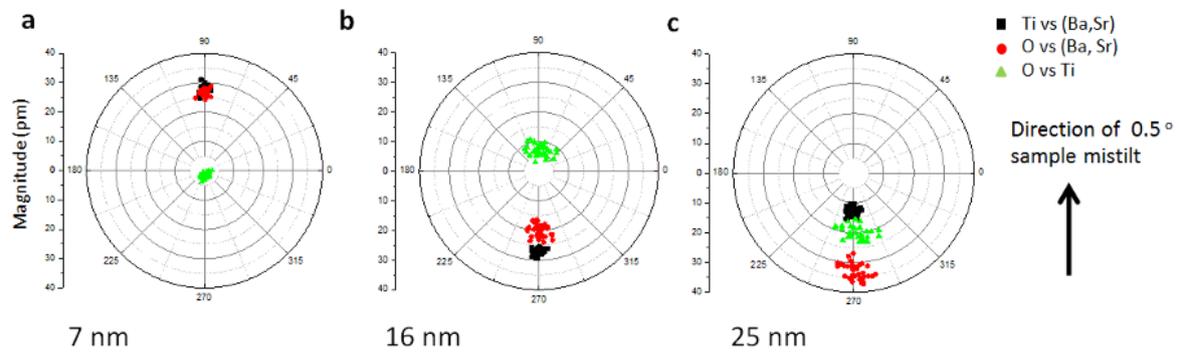

**Figure SI2-4:** Polar figures of Ti vs (Ba, Sr), O vs (Ba, Sr) and O vs Ti displacement measured from ABF images at **a,** 7 nm, **b,** 16 nm and **c,** 25 nm sample thicknesses, at a mistilt angle of 0.5º.

To illustrate that our experimental ABF images are not affected by the specimen mistilt we analysed two regions with different thicknesses and performed displacement measurements as shown in Fig. SI2-5. We can see that we obtain practically identical displacements in magnitude and direction, indicating negligibly sample mistilt. As explained in the main text (see Fig.2) the analysed area can be ascribed to non-cubic symmetry.

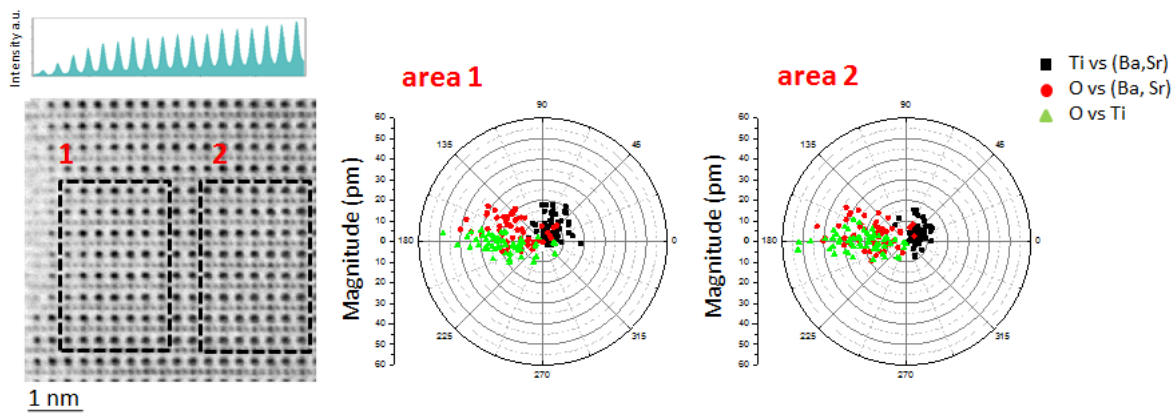

**Figure SI2-5:** ABF image along [110]$_{pc}$ zone axis with displacements of Ti vs (Ba, Sr), O vs (Ba, Sr) and O vs Ti taken from two marked areas with different thickness as indicated by intensity profile taken from corresponding HAADF image shown in Fig.2.

**SI3- BaTiO$_3$ simulations and displacement measurements**

To correlate the experimental atom column displacements with theoretical displacements we simulated ABF images of BaTiO$_3$ as explained for simulations of BST6040 images in SI2. We created a cubic *Pm-3m* (ICSD #27970), orthorhombic *Amm2* (ICSD #161341), tetragonal *P4mm* (ICSD #154343) and rhombohedral *R3m* ((ICSD #73635) structural models of 5 nm thickness. Using these models we then simulated the ABF images in [110]$_{pc}$ zone axis (in pseudo cubic notation) and using 2D Gaussian fit we extracted the exact Ba, Ti and O atom column positions and calculated Ti vs Ba, O vs Ba and O vs Ti displacements. In Fig. SI3-1 the displacements are compiled in the composites polar plots, where the directions of displacements are more clearly delineated.

    The error in the displacements determination is estimated to about 2 pm from calculated displacements in the simulated cubic structure. While in the cubic structure all displacements are by definition zero, the method accurately predicts split of atomic displacements within tetragonal, orthorhombic and rhombohedral phases.

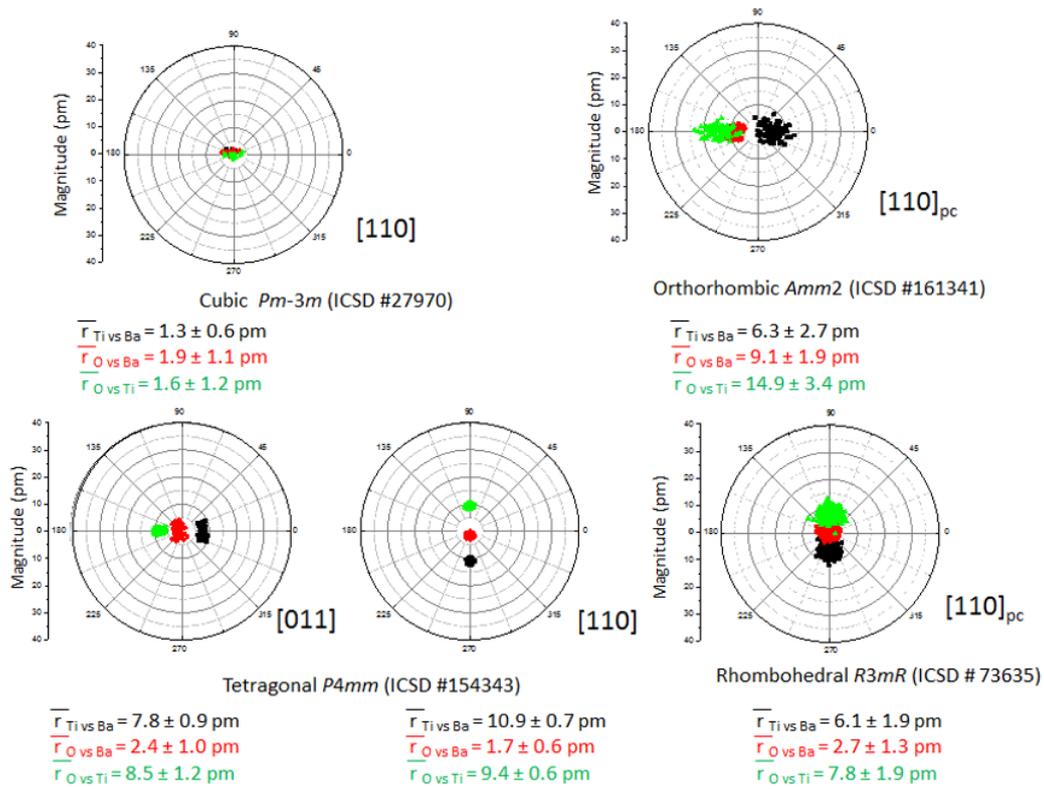

**Figure SI3-1 BaTiO₃ simulations.** Composite pole figures of different displacements for cubic, tetragonal, orthorhombic and rhombohedral symmetries with corresponding average absolute displacement values. As expected in cubic symmetry displacements are around 0. In all other symmetries displacements are split (Ti vs Ba and O vs Ti are in the opposite directions). In tetragonal and rhombohedral symmetries O vs Ba displacements are distributed around origin of the coordinate system, while in the orthorhombic symmetry O vs Ba displacements are centred away from origin of the coordinate system.

Experimental atomic displacements were first measured on BaTiO₃ sample at room temperature (i.e., in the tetragonal phase) along $[110]_{pc}$ zone axis, Fig. SI3-2. The spread of atomic shits is more than twice larger in the experimental than in modelled images; the maximal standard deviation in the experimental case is 7 pm versus 3 pm in the modelled structure indicating disorder in the sample (compare Fig SI3-1, Fig SI3-2). This disorder could be indicative of a non-negligible concentration of cationic and O vacancies, whose presence would distort the structure. Alternatively, it could be a consequence of a ferroelectric domain wall within the examined area.

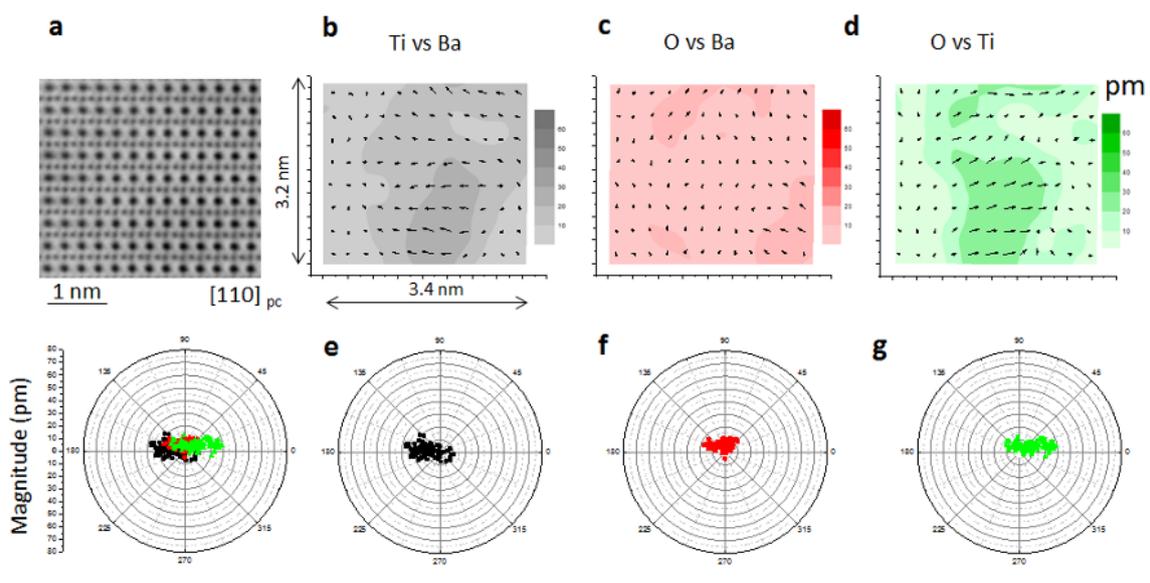

**Figure SI3-2 BaTiO$_3$ experimental results at RT before *in situ* heating**. **a**, ABF image along [110]$_{pc}$ zone axis taken at RT (before *in situ* experiment) with a composite pole plot; **b-d,** maps of displacements (arrows represent the direction of the displacements and their magnitude in pm- see colour scale on the right) with corresponding **e-f** pole figures. The average values of the relative atomic displacements measured over the whole image are 12 ± 6 pm for Ti vs Ba, 8 ± 4 pm for O vs Ba, and 13 ± 7 pm for O vs Ti. Due to antiparallel displacement of Ti vs Ba compared to O vs Ti and displacements of O vs Ba around centre, the overall structure seems, as expected, to be tetragonal.

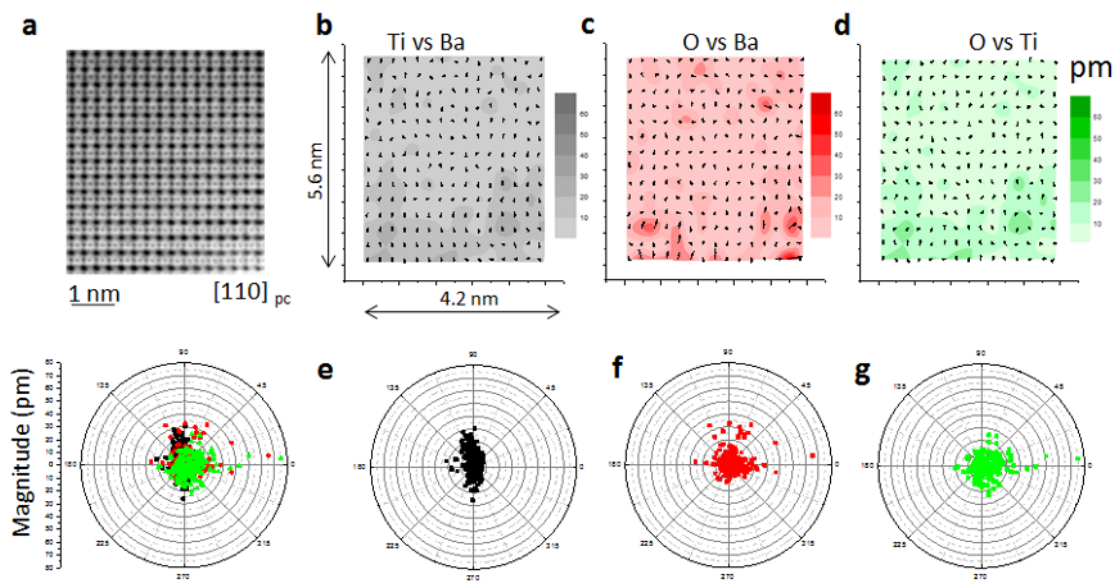

**Figure SI3-3 BaTiO$_3$ relative atomic displacements at 300°C. a,** ABF image along [110]$_{pc}$ zone axis taken at 300°C with a composite pole plot. **b-d,** Maps of displacements (arrows represent the direction of the displacements and their magnitude in pm- see colour scale on the right) with corresponding **e-f** pole figures. The average values of the relative atomic displacements measured over the whole image are 10 ± 6 pm for Ti vs Ba, 10 ± 8 pm for O vs Ba, and 11 ± 9 pm for O vs Ti.

Splitting of displacements is much smaller than at 200°C indicating that symmetry is approaching cubic.

**SI4 Quantitative HAADF STEM analysis of atom column intensities and average chemical composition of (Ba, Sr)TiO$_3$**

To determine local stoichiometry of BST6040, normalized (Ba, Sr)-column intensities were measured using the approach by LeBeau & Stemmer[12] where the intensity of the background and individual atom columns were measured from the HAADF images at non-saturating settings of brightness and contrast. The detector background intensity was subtracted from the intensity of each pixel in experimental images. (Ba, Sr) and Ti column intensities were extracted form HAADF images as intensity peak integral, integrated within roughly two sigma by approximating a 2D Gaussian-type peak. The final values of individual atom-column intensities were normalised to the highest column intensity in the analysed area to cancel the influence of HAADF detector amplification (for details see also ref. [13]). A quantitative estimate of the intensity distribution of A sites due to chemical fluctuations in BST6040 was then obtained by simulating HAADF image of a cubic structural model of 10 nm thickness in [110] zone axis image (for simulation details see SI2) with randomly distributed Ba and Sr atoms in 60/40 ratio. For each A site, using random generator, Ba or Sr atom were assigned in 0.6/0.4 probability ratio.

We get from the model in [110]$_{pc}$ orientation that the expected average intensity for a random distribution of Ba$_{0.60}$ and Sr$_{0.40}$ on A-site is 0.88±0.05. The intensity is expressed in relative terms, with intensity 1 being assigned to a column with the highest intensity. The average experimental intensity value obtained is 0.94±0.02 for a 20 nm thick BST6040 sample (Fig. SI4-1), and 0.90±0.02 for a 5 nm thick sample (not shown). These values are within the experimental error of a perfectly random distribution of Br and Sr with 60/40 stoichiometry.

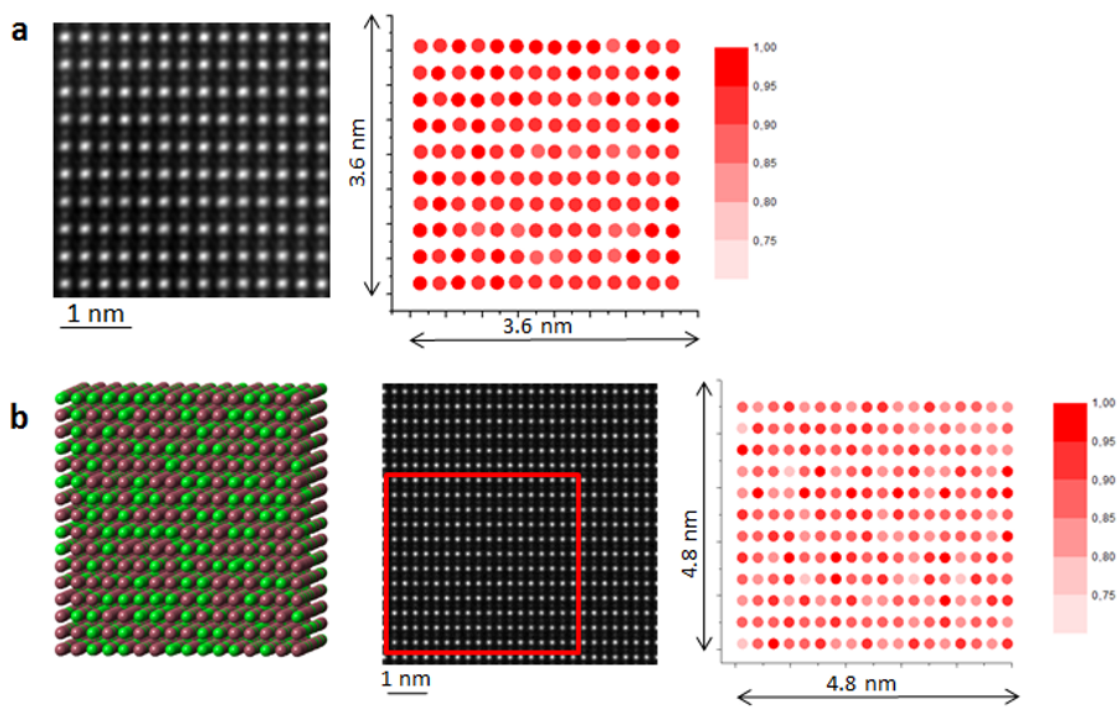

**Figure SI4-1. Average chemical composition. a,** Experimental HAADF image of BST (simultaneously taken with ABF image shown in Figure 2 in [110]$_{pc}$ zone axis with the corresponding normalized intensity distribution map of (Ba, Sr)-columns. Average relative intensity of (Ba, Sr) column intensity is 0.93 ± 0.02. **b,** A part of the model showing only Ba (brown) and Sr (green) atoms, with the simulated HAADF image and corresponding normalized intensity distribution map of (Ba, Sr) columns. Chemical homogeneity (average relative intensity of Ba, Sr-site atom columns) estimated for a 15 nm thick BST6040 model where Ba and Sr randomly occupy A sites is 0.88 ± 0.05. The results indicate that there is no significant departure in the experimental data from the random distribution of Ba and Sr atoms with 60/40 stoichiometry.

## SI5 Influence of overlapping polar clusters along the viewing direction on displacement measurements

To estimate the effects of embedded clusters on displacement measurements, two situations were considered: i) a polar tetragonal volume embedded into a cubic matrix (Fig. SI5-1), and (ii) four alternating polar and cubic (nonpolar) regions (Fig. SI5-2).

In the first model, a 4 nm-thick stack of BST6040 with tetragonal structure was embedded into a cubic volume, of the same composition, Fig. SI5-1. Simulated ABF image of the model in [110] zone axis gives O vs (Ba,Sr) and O vs Ti displacements of about 12.5 ± 1.1 pm, which is close (within experimental accuracy) to the assigned displacements of 14.7 pm in the modelled tetragonal BST6040.

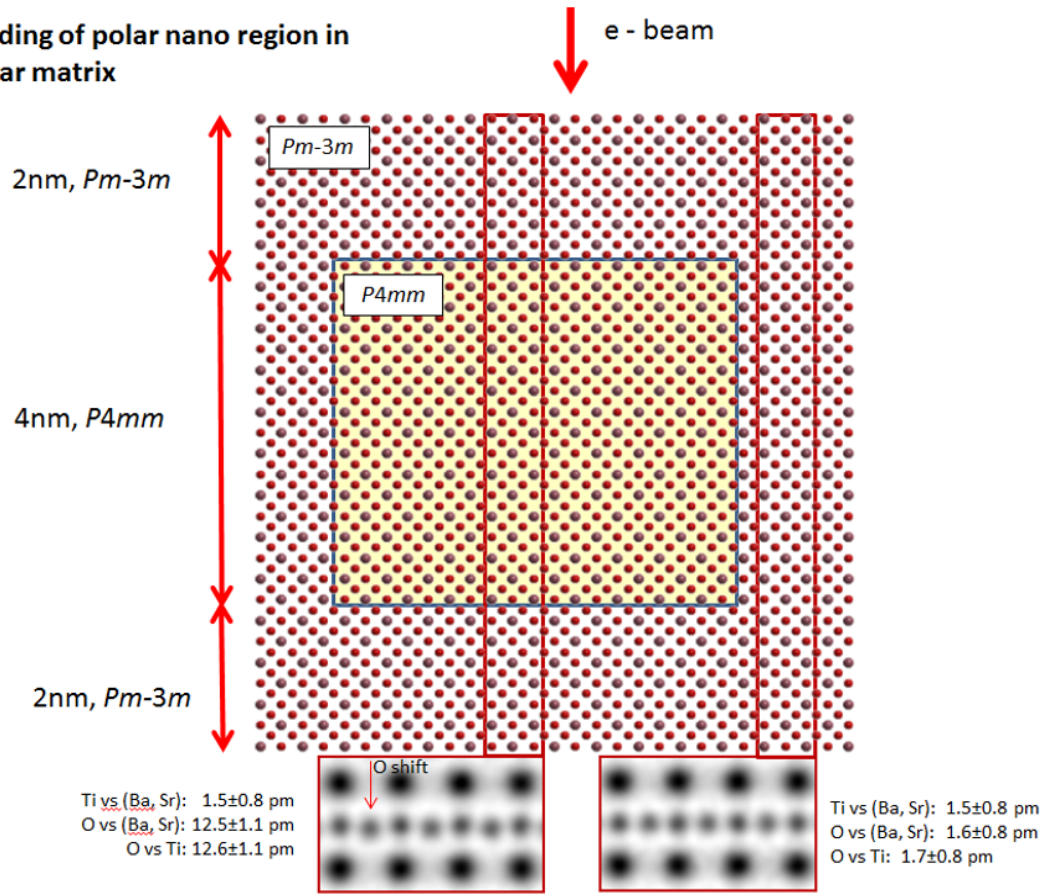

**Figure SI5-1 Simulated displacements in BST6040.** Schematic of simulated tetragonal *P4mm* (ICSD # 188786) embedded volume in cubic *Pm-3m* (ICSD # 90006) matrix (with indicated direction of the electron beam (e-beam) and dimensions) and ABF images calculated from the model in overlapped (left) and pure cubic volumes (right). The displacements set for the tetragonal volume, calculated from the *P4mm* model are Ti vs (Ba, Sr) 0 pm, O vs (Ba, Sr) 14.7 pm and O vs Ti 14.7 pm. Measured displacements in overlapped volume are Ti vs (Ba, Sr) 1.5±0.8 pm, O vs (Ba, Sr) 12.5±1.1 pm and O vs Ti 12.6±1.1 pm, that is all within experimental error. In pure cubic phase displacements are, as expected, close to 0, i.e. Ti vs (Ba, Sr) 1.5±0.8 pm, O vs (Ba, Sr) 1.6±0.8 and O vs Ti 1.7±0.8 pm. Based on these results we estimate that the first up to ~6 nm-thick surface region (where the beam is entering) crucially influences the image formation and consequently the displacement measurements.

In the second situation, with alternating polar/nonpolar layers, we considered three cases, as indicated in Fig. SI5-2. In Cases 1 and 2, ABF images were simulated from the entire, 24 nm-thick volume, but with opposite electron beam direction, as indicated in the Figure. In Case 3, ABF image was simulated only from the top 3 nm layer (see Figure SI5-2). For the three cases the average displacements in pm are collected in the corresponding Table. If all parts of the sample in Cases 1 and 2 would contribute equally to the image we should get all values around 0, but this is not the case. For Ti vs (Ba, Sr), in Case 1, we can state that there are no displacements, if we ascribe the values of ~6 pm to the experimental error. O vs (Ba, Sr) and O vs Ti are almost doubled compared with displacement in individual tetragonal region (~12 pm, see Case 3). Obviously, opposite polarization does not cancel out; the main information is coming from the top region of the sample. The same conclusion can be made from Case 2, where all displacements are close to zero, because the cubic phase, where electrons enter the sample, prevails. In Case 3, where the sample is thin and has tetragonal symmetry, we measured the expected values (~0 for Ti vs (Ba, Sr) and ~12 pm for O vs (Ba, Sr) and O vs Ti).

Top few nm of the sample (where the beam is entering) has the largest effect on the image and consequently on the measured displacements. So even in thicker samples (20 nm) we may determine the presence of polar nano-regions if they are at the top part of the sample. In thicker samples the measured displacements are overestimated but their direction is correctly determined.

To summarize, based on both results (Fig. SI5-1 and SI5-2) we can estimate that the first 4-5 nm thick surface region (where the beam is entering) crucially influences the image formation and consequently the displacement measurements.

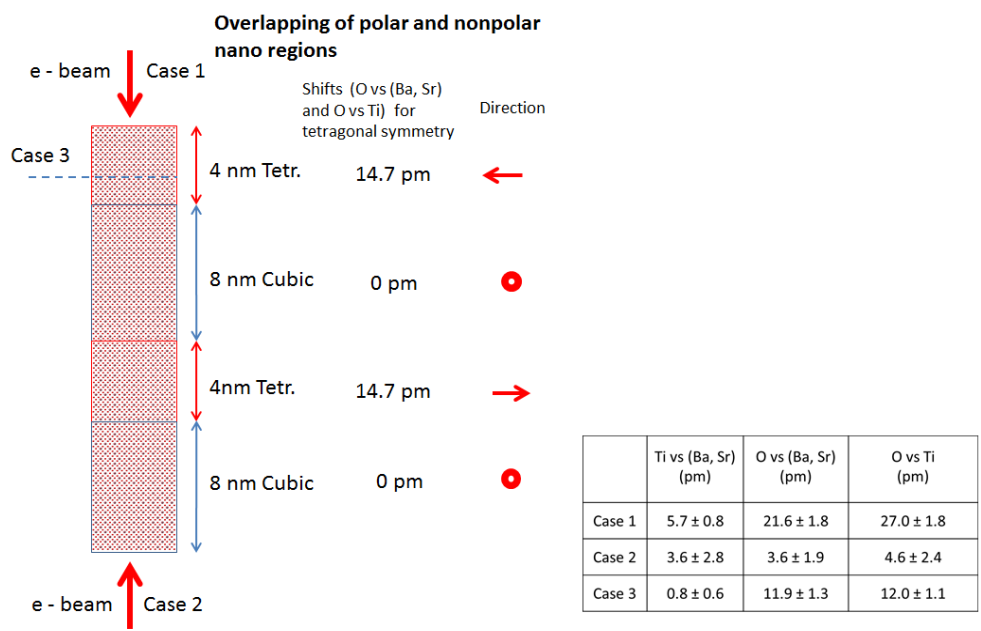

**Figure SI5-2 Influence of thickness and overlapping layers on displacement determination.** A model was constructed where 4 layers of alternating tetragonal and cubic phases were placed across a 24 nm-thick sample. The polarizations of two tetragonal areas were in opposite directions. Based on the model, three different ABF images were simulated. In Case 1 the electron beam was entering the sample from top to bottom through the whole thickness (24 nm). In Case 2 the electron beam was entering from bottom to top through the whole thickness and in the Case 3 the electron beam was entering as in Case 1 but the image was created just from the first 3 nm-thick volume.

## SI6: Evidence of strain associated with polar nanoclusters

We present two independent results, Fig. SI6-1 and SI6-2, as evidence for strain associated with polar nanoclusters.

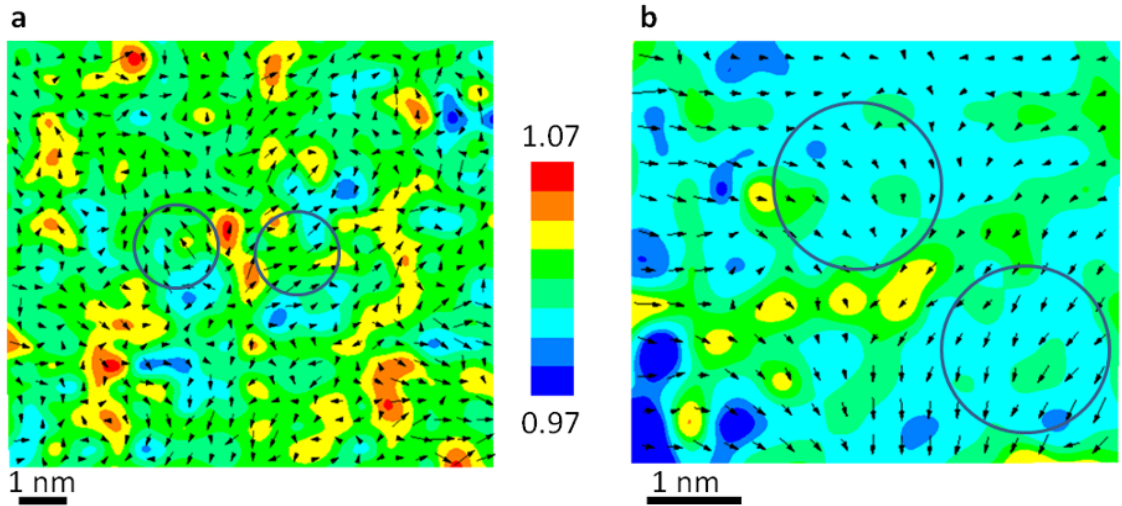

**Figure SI6-1. Lattice strain distribution. a,** Lattice-strain map of BST6040 along $[001]_{pc}$ zone axis (from HAADF image in Fig. 1**a**). The lattice strain distribution was determined by extracting the horizontal (x) and vertical (y) (Ba, Sr) lattice parameters and is represented by the x/y ratio colour map. The map of the x/y ratio is superimposed on corresponding Ti vs (Ba, Sr) displacements, represented by orientation and length of the associated arrows. A correlation between the change of displacement direction and strain magnitude is visible. Maximal strains are observed at the interfaces between polar regions (as indicating by circles), suggesting that each region with coherent displacement of atoms is associated with specific strain distribution; **b,** Lattice-strain map of $BaTiO_3$ along $[110]_{pc}$ zone axis (From ABF image in Fig. 4**a**), The map of x/y ratio is superimposed on corresponding Ti vs Ba displacements at 200°C. A correlation between change of displacement direction and strain magnitude can be noticed. Polar regions are indicated by circles.

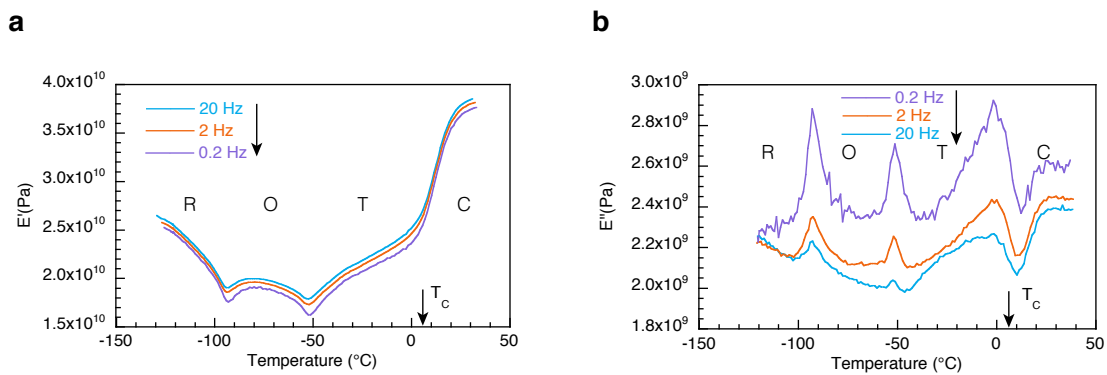

**Figure SI6-2. Elastic modulus E as a function of temperature for BST6040. a,** Real and **b,** Imaginary component of E. The C, T, O and R mark cubic, tetragonal, orthorhombic and rhombohedral phase regions, respectively. Compare with Fig. SI1-1**b**. Two features are indicative of presence of elastically active nanosize objects within the material: (i) the large decrease of the elastic modulus E' in the cubic phase as the temperature is decreased toward Curie temperature $T_C$

and (ii) the frequency dependence of E' and E" in the cubic phase.[14,15] In the ferroelectric phases the temperature dependence and relaxation of E' and E" are dominated by ferroelastic domains. These results corroborate interpretation of microscopic strain data shown in Fig. SI6-1.

The complex elastic modulus E (mechanical stiffness or storage modulus and loss) was measured in the single cantilever mode with a Perkin- Elmer PYRIS Diamond Dynamic Mechanical Analyzer, with the temperature rate of 1–2 K/min.